\documentclass[
aip, amsmath,amssymb, reprint,
]{revtex4-1}

\usepackage{graphicx}
\usepackage{dcolumn}
\usepackage{bm}

\usepackage[utf8]{inputenc}
\usepackage[T1]{fontenc}
\usepackage{mathptmx}
\usepackage{etoolbox}
\usepackage{physics}

\makeatletter
\def\@email#1#2{%
 \endgroup
 \patchcmd{\titleblock@produce}
  {\frontmatter@RRAPformat}
  {\frontmatter@RRAPformat{\produce@RRAP{*#1\href{mailto:#2}{#2}}}\frontmatter@RRAPformat}
  {}{}
}%
\makeatother
\begin{document}

\preprint{AIP/123-QED}

\title{Designer gapped and tilted Dirac cones in lateral graphene superlattices}

\author{A. Wild}
\affiliation{%
 Physics and Astronomy, University of Exeter, Stocker Road, Exeter EX4 4QL, United Kingdom
}%
\author{R. R. Hartmann}
\affiliation{%
 Physics Department, De La Salle University, 2401 Taft Avenue, 0922 Manila, Philippines
}%
\author{E. Mariani}
\affiliation{%
 Physics and Astronomy, University of Exeter, Stocker Road, Exeter EX4 4QL, United Kingdom
}%

\author{M. E. Portnoi}
\email{m.e.portnoi@exeter.ac.uk}
\affiliation{%
 Physics and Astronomy, University of Exeter, Stocker Road, Exeter EX4 4QL, United Kingdom
}%

\date{\today}

\begin{abstract}
We show that a planar array of bipolar waveguides in graphene can be used to engineer gapped and tilted two-dimensional Dirac cones within the electronic band structure. The presence of these gapped and tilted Dirac cones is demonstrated through a superlattice tight-binding model and verified using a transfer matrix calculation. By varying the applied gate voltages, the tilt parameter of these Dirac cones can be controlled and their gaps can be tuned to fall in the terahertz range. The possibility of gate-tunable gapped Dirac cones gives rise to terahertz applications via interband transitions and designer Landau level spectra both of which can be controlled via Dirac cone engineering. We anticipate that our paper will encourage Dirac cone tilt and gap engineering for gate-tunable device applications in lateral graphene superlattices.
\end{abstract}

\maketitle

\section{Introduction}
The relativistic nature of graphene's charge carriers leads to its fascinating optical and electronic properties~\cite{RevModPhys.81.109,doi:10.1126/science.1156965}. Its discovery opened the door to the exploration of relativistic physics in condensed matter systems. Indeed, the rise of graphene inspired the search for new designer materials with ultra-relativistic spectra, such as 8-$Pmmn$ borophene. This theoretical material is predicted to contain two-dimensional (2D) tilted Dirac cones in its electronic band structure in the vicinity of the Fermi level~\cite{PhysRevLett.112.085502}. With its discovery came an explosion of interest into the physics arising from tilted Dirac cones. These cones can either be gapped or gapless, and come in three types: type-I (sub-critically tilted), type-II (super-critically tilted) or type-III (critically tilted)~\cite{10.1038_nature15768,PhysRevX.9.031010}. Each geometry giving rise to spectacularly different optical~\cite{doi:10.1143/JPSJ.79.114715,PhysRevB.96.155418,PhysRevB.103.165415,PhysRevB.103.125425,wild2021optical}, transport~\cite{C7CP03736H,PhysRevB.97.235113,PhysRevB.97.235440} and thermal properties~\cite{Sengupta_2018,PhysRevB.99.235413,PhysRevB.102.045417} and more~\cite{doi:10.1143/JPSJ.77.064718,PhysRevB.91.195413,PhysRevB.96.235405,PhysRevB.98.195415,PhysRevB.99.035415,ng2021mapping,Liu_2023,PhysRevB.105.L201408}. For device applications, it would be highly desirable to be able to switch between different types of tilted Dirac cones in a single system post-fabrication.

Currently, there is a dearth of practical, tunable, electronic systems which exhibit 2D tilted Dirac cones. Several theoretical materials with specific lattice geometries have been predicted to support electronic tilted Dirac cones~\cite{PhysRevLett.112.085502,doi:10.1143/JPSJ.75.054705,PhysRevB.78.045415,PhysRevLett.105.037203,doi:10.1126/science.1256815,PhysRevX.6.041069,PhysRevB.94.195423,Ma2016,PhysRevB.95.035151,PhysRevB.95.245421,https://doi.org/10.1002/pssr.201800081,PhysRevB.98.121102,PhysRevB.100.235401,PhysRevB.100.205102,PhysRevB.102.041109}. However, after synthesis, crystalline structures cannot be practically changed to tune the tilt or modify the gap of these cones. Rather than placing real atoms in a particular lattice configuration, we propose to approach the problem using artificial atoms; namely, bound states trapped inside graphene wells and barriers organized in a lateral superlattice.

In contrast to non-relativistic systems, both electrostatic wells and barriers in graphene support bound states. These bound states are localized about the center of the confining potentials, much like atomic orbitals in a crystal are centered about their lattice positions. The confined states of a well and barrier overlap, much like adjacent atomic orbitals. This overlap can be characterized by the hopping parameter in the famous tight-binding model. Unlike a real crystal, where the overlap between adjacent orbitals is fixed, the overlap between well and barrier functions can be completely controlled. This can be achieved by varying the height and depth of the confining potentials via their top-gate voltages. Hence, constructing a superlattice from wells and barriers in graphene mimics the band structure of an atomic lattice but with the advantage of a newfound tunability. Thus, moving band structure engineering in condensed matter physics in the same direction as optical control in designer metamaterials\,\cite{10.1038_nmat3839}.

\begin{figure}
    \centering
    \includegraphics[width=0.35\textwidth]{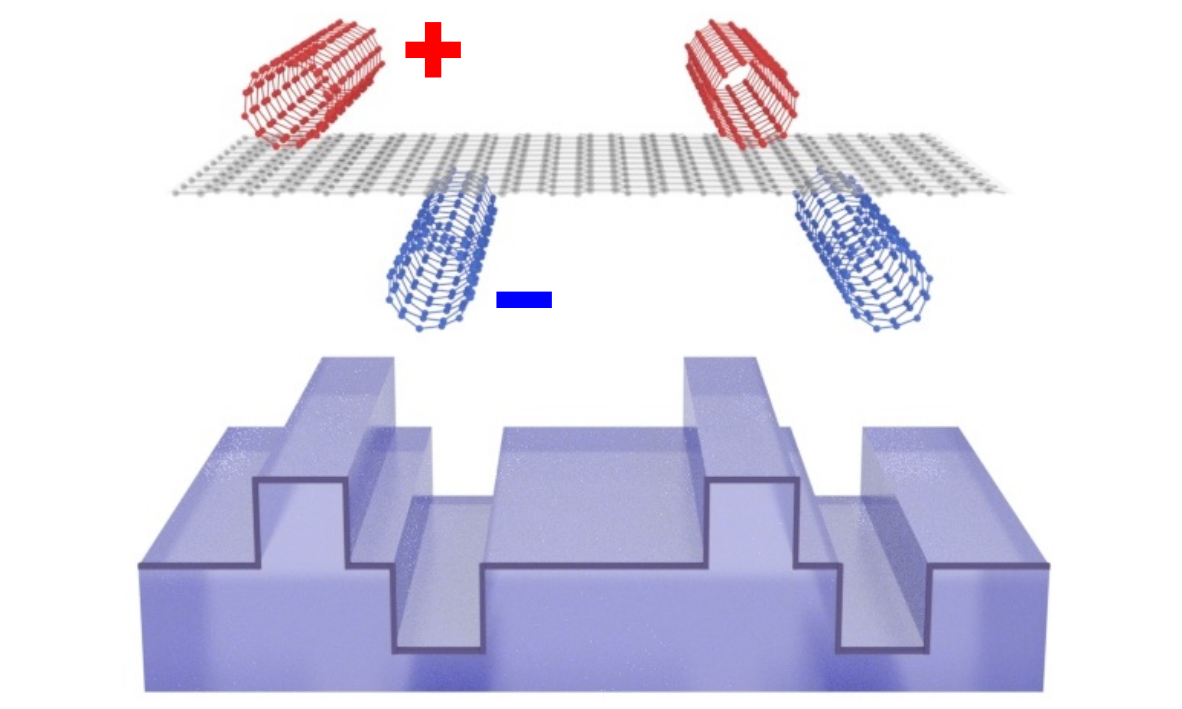}
    \caption{Schematic of a planar array of bipolar waveguides in graphene created by carbon nanotubes gated with alternating polarity. The electrostatic potential created by the applied gate voltages is shown below.}
    \label{fig:Schematics}
\end{figure}

In what follows, we show that a lateral superlattice comprised of repeating well and barrier pairs, i.e., bipolar array (see Fig.\,\ref{fig:Schematics}), hosts gapped and tilted Dirac cones in the band structure. By varying the applied voltage profile of the superlattice, the tilt of these Dirac cones can be controlled and the band gap can be tuned to energies corresponding to terahertz (THz) photons. By calculating the velocity matrix element in the vicinity of the gapped Dirac cones we prove that the bipolar array in graphene will constitute a platform for tunable terahertz optics. Whilst we demonstrate the existence of tunable tilted and gapped Dirac cones in the electronic band structure, we emphasize that these cones are satellites to a central gapless cone. In contrast to Dirac cones in pristine graphene, these central cones are anisotropic in momentum space, possessing elliptical isoenergy contours. Applying a magnetic field normal to the plane of the bipolar array creates a platform for a gate-tunable Landau level spectra via Dirac cone engineering. Due to the presence of central gapless and satellite gapped Dirac cones, the Landau level spectra simultaneously contains features of massless and massive Dirac fermions.

\section{Model}
A lateral graphene superlattice can be modeled as an artificial crystal. Whilst in a crystalline material, electrons hop between adjacent atomic orbitals, in a lateral superlattice, electrons hop between neighboring well and barrier sites. Thus, to calculate the band structure of a bipolar array in graphene we shall use a simple nearest-neighbor tight-binding model.

\begin{figure}
    \centering
    \includegraphics[width=0.44\textwidth]{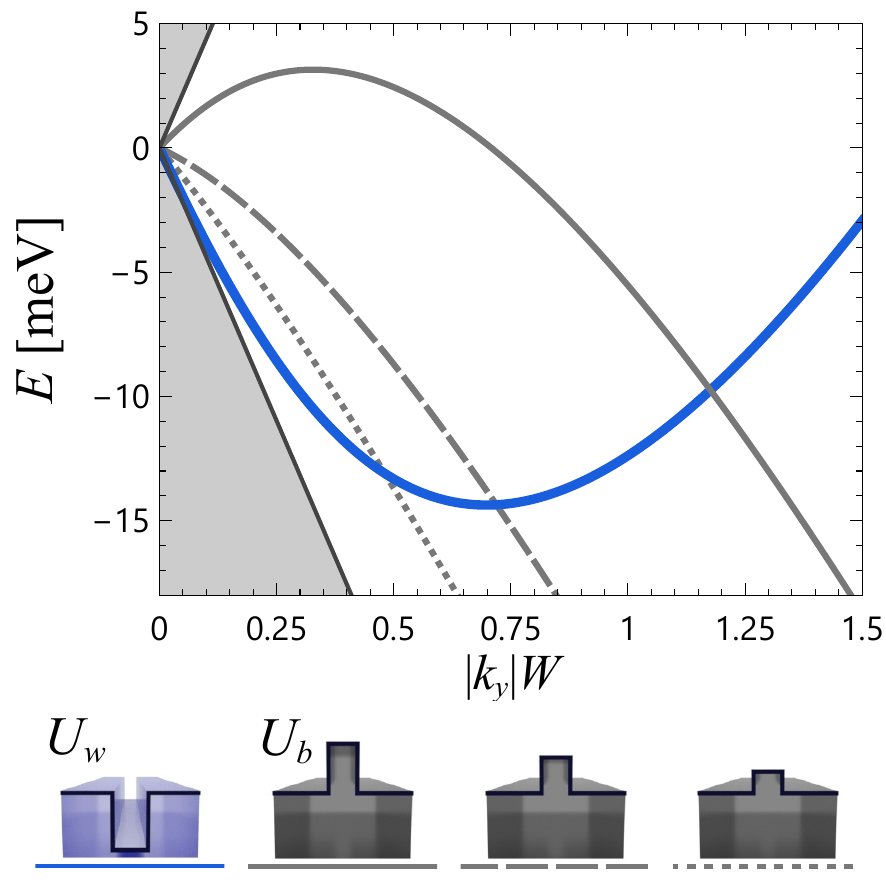}
    \caption{The energy spectrum of confined states within a well of applied voltage $U = -120\text{meV}$ (blue) and three barriers of strengths $90\text{meV}$ (grey solid), $60\text{meV}$ (grey dashed) and $45\text{meV}$ (grey dotted) in graphene - in each case the well/barrier width is $W = 15\text{nm}$. The band dispersions are sketched in units of energy $E$ in milielectron volts (meV) and wavevector associated to motion along the potentials $k_y$ normalized by the well/barrier width. The formed band crossings are of type-I, -III and -II respectively. Tuning the barrier height and well depth changes the tilt of the band crossings. The grey regions contain continuum states outside of the guiding potentials.}
    \label{fig:WellBarrierSchem}
\end{figure}

Let us first consider one artificial atom (i.e., a square quantum well or barrier) in our superlattice. Although realistic top-gated structures in graphene generate smooth guiding potentials\,\cite{PhysRevA.89.012101,PhysRevB.81.245431,10.1038/s41598-017-11411-w} (i.e., varying on a length scale much larger than the lattice constant) they can be modeled as square potentials with an effective depth and width. This is because the number of bound states in a potential is dictated by the product of its effective depth and width. It should be noted that in what follows we consider sharp-but-smooth square potentials, i.e., we neglect inter-valley scattering. The effective one-dimensional matrix Hamiltonian for confined modes in a graphene waveguide can be written as
\begin{equation}
\label{eq:DiracEffective}
    \big[ \hat{H}_\text{G} + U(x)\mathbb{I} \big]
    \ket{\psi(x)} = E\ket{\psi(x)},
\end{equation}
where $\hat{H}_\text{G} = v_\text{F} ( \sigma_x \hat{p}_x + s_\text{K} \sigma_y \hbar k_y )$, which acts on the spinor wavefunction defined in the standard basis of graphene sub-lattice Bloch sums $\ket{\psi(x)} = \psi_A(x)\ket{\varPhi_A} + \psi_B(x)\ket{\varPhi_B}$. The Pauli matrices are $\boldsymbol{\sigma} = (\sigma_x,\sigma_y,\sigma_z)$, the identity matrix is $\mathbb{I}$, the momentum operator is defined as $\hat{p}_{x} = -i \hbar \partial_{x}$, and $k_y$ is a wavenumber corresponding to the motion along the waveguide. Here, the Fermi velocity in graphene is $v_\text{F} \approx 10^6 \text{ms$^{-1}$}$, the energy eigenvalue is $E$, and the graphene valley index is $s_\text{K} = \pm 1$. The square potential $U(x)$ is defined as
\begin{equation}
    U(x) = \begin{cases}
    U
    & \mid \! x \!\mid \: \leq W/2\\
    0, & \text{elsewhere}
    \end{cases},
    \label{eq:potentials}
\end{equation}
where $W$ is the width of the potential and $U < 0$ for a well and $U>0$ for a barrier. The eigenvalues of Eq.~(\ref{eq:DiracEffective}) can be obtained via the method outlined in Ref.~\cite{PhysRevB.74.045424}. For zero-energy states ($E = 0$), the eigenvalue problem simplifies and the wavefunction takes on a simple form (see Appendix~\ref{sec:ZeroEnergyStates}) - these zero-energy well and barrier wavefunctions will be utilized later in the paper. In Fig.~\ref{fig:WellBarrierSchem} we superimpose the energy spectra for various square wells and barriers. Each potential has the same width and contains only a few modes within. This occurs when the normalized product of the potential height and width ($\left| U \right| \! W/\hbar v_\text{F}$) is of the order of unity. Indeed, few-mode smooth electron waveguides in graphene can be experimentally realized using carbon nanotubes as top-gates\,\cite{PhysRevLett.123.216804}.

\begin{figure*}
    \centering
    \includegraphics[width=0.85\textwidth]{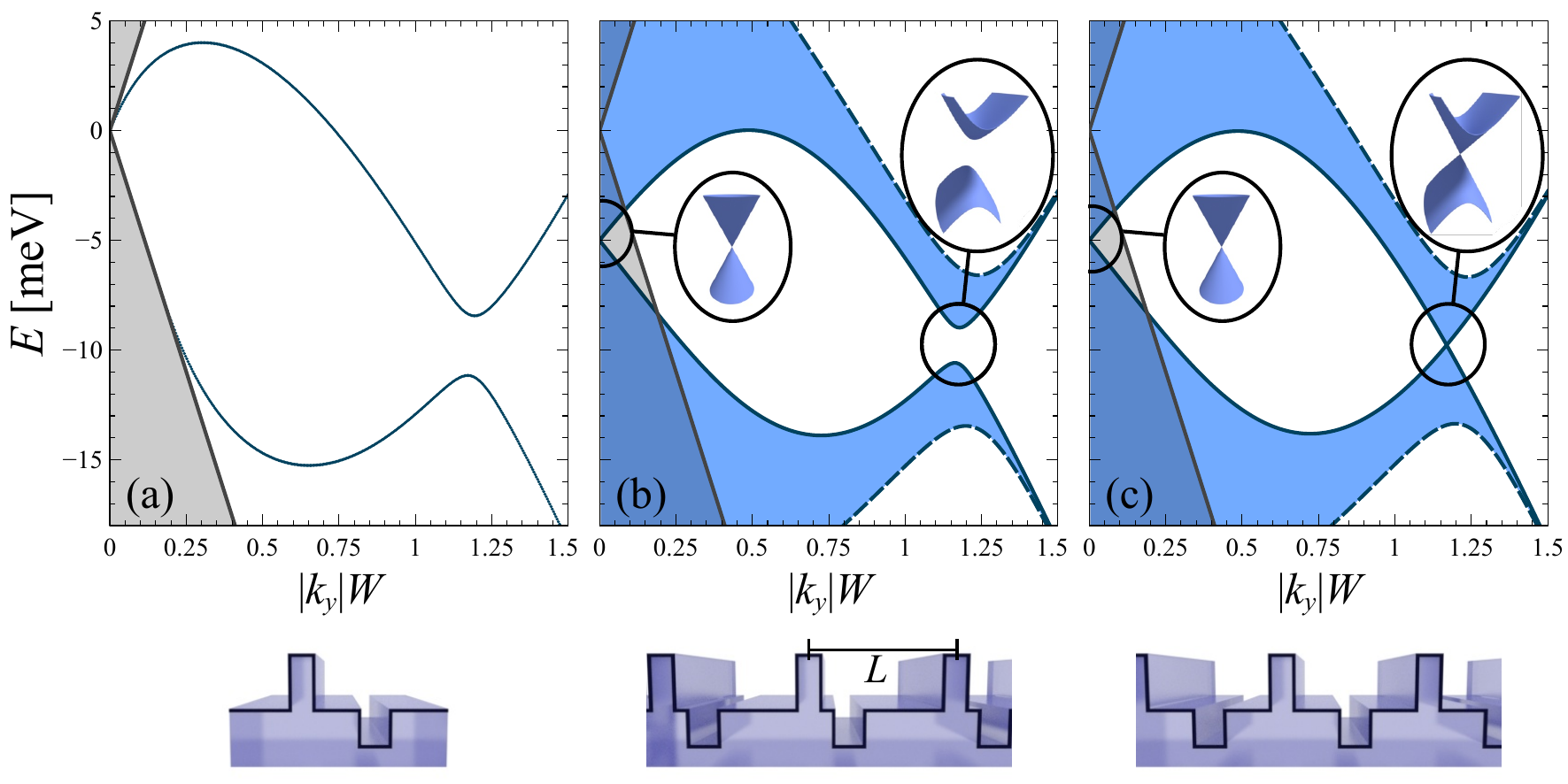}
    \caption{The band structure and schematics of (a) single bipolar waveguide, (b) bipolar array without reflection symmetry, (c) bipolar array with reflection symmetry. The bipolar array with or without reflection symmetry possesses a gapless Dirac cone at the center of the electronic band structure. Additionally, the bipolar array without reflection symmetry hosts satellite gapped tilted Dirac cones whilst the bipolar array with reflection symmetry hosts satellite gapless tilted Dirac cones. The band structures are plotted in terms of energy ($E$) in units of milielectron volts (meV) and wavevector along the potentials $k_y$ normalized by the well/barrier widths $W = 15\text{nm}$. In all cases the applied barrier and well potentials are $U_b = 90\text{meV}$ and $U_w = -120\text{meV}$ respectively. For the bipolar array, the unit cell width is $L = 90\text{nm}$, the position of the well within the unit cell is determined by the parameter $a = 48\text{nm}$ in panel (b) and $a = 45\text{nm}$ in panel (c). The grey areas correspond to energies and wavevectors which support plane wave solutions across the entire potential. The periodicity of the superlattice yields an additional wavevector $\left| k_x \right| \leq \pi/L$ where $L$ is the size of the unit cell. The band structures in panels (b) and (c) were calculated using a transfer matrix model. These panels display orthographic projections of the band structures as viewed along the $k_x$ axis. The band edges are depicted by solid lines ($k_x = 0$) or dashed lines ($k_x = \pm \pi/L$) with intermediate values shaded in blue. 
    }
    \label{fig:BipolarToArray}
\end{figure*}

As can be seen from Fig.\,\ref{fig:WellBarrierSchem}, varying the potential strengths of the well and barrier results in differing group velocities at the crossing point. Thus, the crossing formed by an isolated well and barrier can be switched between type-I, type-II or type-III by simply changing the potential strength of the well and barrier. However, by superimposing the band dispersions of an isolated well and barrier we have neglected any coupling between the two systems. When we bring the well and barrier closer together the overlap between barrier and well states leads to an anticrossing (pseudogap) appearing at the original band crossing (see Fig.~\ref{fig:BipolarToArray}(a)) ~\cite{PhysRevB.102.155421}. For bipolar waveguides created by carbon nanotube top-gates atop graphene, the pseudogap is of the order of several THz~\cite{PhysRevB.102.155421,PhysRevA.102.052229}. However, a single well and barrier does not constitute a macroscopic device, and for realistic THz applications an important question must be answered: How is the band structure of a single bipolar waveguide modified when placed in to a superlattice?

The Hamiltonian of a planar bipolar array $\hat{H}$ can be written as
\begin{equation}
\label{eq:SuperlatticeHamiltonian}
\hat{H}=\hat{H}_{\text{G}}+\sum_{j=1}^{N}\big[ U_{b}\left(x-x_j\right)+U_{w}\left(x-x_{j}-a\right) \big],
\end{equation}
where $U_b(x)$ and $U_w(x)$ are individual barriers and wells (defined through Eq.~(\ref{eq:potentials})), centered at positions $x_{j}$ and $x_{j}+a$, respectively, where $x_j = jL$ is a lattice vector, $L$ is the width of the superlattice unit cell and $a$ is the distance between the centres of a well and barrier within the unit cell. Since the dispersion of any realistic guiding potential is determined by the product of the potential's depth and width, henceforth we fix the width of all barriers and wells $W$ to be the same. For square potentials, the well position within the unit cell must satisfy $a > W$ whilst the unit cell width must be larger than the sum of the well and barrier width $L > 2W$. 

One may envisage a bipolar array created by sandwiching a graphene sheet in between two planar arrays of nanotubes, with the top array gated at one polarity, the bottom array at the opposite polarity. The relative position of these two arrays (parameterized by $a$) will be fixed after device fabrication. In a realistic device, it will not be possible to exactly align the two arrays such a way that each tube is equally separated, in general the two arrays will be separated by some arbitrary distance ($a \neq L/2$).

\subsection*{Tight-binding model of a bipolar array in graphene}
In a similar fashion to the splitting of atomic energy levels in the formation of a crystal, the bringing together of $N$ bipolar waveguides results in each energy level of the well and barrier splitting in to $N$ sub-levels. Each sub-level corresponds to a particular quantized $k_x$. In the limit that $N$ becomes large, $k_x$ can be treated as a continuous parameter on an equal footing with $k_y$, the wavevector along the guiding potentials.

The basis functions of the superlattice can be expressed as a linear combination of individual well and barrier wave functions, i.e., Bloch sums:
\begin{equation}
\label{eq:BlochSum1}
    \ket{\Phi_b} = \frac{1}{\sqrt{N}} \sum_{j=1}^N e^{i j k_{x} L} \ket{\psi_{b}(x - x_j)},
\end{equation}
\begin{equation}
\label{eq:BlochSum2}
    \ket{\Phi_w} = \frac{e^{i k_x a}}{\sqrt{N}} \sum_{j=1}^N e^{i j k_{x} L} \ket{\psi_{w}(x - x_{j} - a)},
\end{equation}
where $\left| k_x \right| \leq \pi/L$ is the superlattice wavevector and the well and barrier functions $\ket{\psi_w(x)}$ and $\ket{\psi_b(x)}$ are the solutions to Eq.~(\ref{eq:DiracEffective}) with potentials $U_w(x)$ and $U_b(x)$ respectively for a given wavevector along the guiding potentials $k_y$. Here, we have utilized the so-called atom gauge where the orbital centers of the well and barrier states are encoded in the phase of the Bloch sums\,\cite{10.21468/SciPostPhysCore.6.1.002} (another common choice is the cell gauge where the $e^{i k_x a}$ term is omitted from Eq.\,(\ref{eq:BlochSum2})). The eigenvalues of the superlattice as a function of $k_x$ (the superlattice wavevector) are determined from the secular equation $\text{det}(\mathcal{H} - E\mathbb{I}) = 0$, where the elements of the Bloch Hamiltonian are defined as $\mathcal{H}_{\alpha\beta} = \int \bra{\Phi_\alpha}\hat{H}\ket{\Phi_{\beta}}\text{d} x$, where $\alpha,\beta=w$ or $b$. 

We intend to model the electronic dispersion of a superlattice in the vicinity of the original band crossings of the first well and barrier modes (see Fig.~\ref{fig:WellBarrierSchem}). These crossings occur at wavevectors $k_y = K_y = s \! \left| K_y \right|$ where $s = 1$ or $-1$. To determine the diagonal elements of the Bloch Hamiltonian we approximate the well and barrier band dispersion with linear functions, i.e.,
\begin{equation}
\label{eq:vbarr}
    \int_{-\infty}^\infty \bra{\psi_b(x)} \hat{H} \ket{\psi_b(x)}\text{d} x \approx s \hbar v_b (k_y-K_y),
\end{equation}
and
\begin{equation}
\label{eq:vwell}
    \int_{-\infty}^\infty \bra{\psi_w(x)} \hat{H} \ket{\psi_w(x)}\text{d} x \approx s \hbar v_w (k_y-K_y),
\end{equation}
where $v_w > 0$ and $v_b < 0$ are the well and barrier group velocities at the crossing point (see Fig.~\ref{fig:WellBarrierSchem}). The inclusion of higher-order terms such as an effective mass is discussed in Section\,\ref{sec:Tilt}. Note that the offset energy ($E_\text{off}$) of these functions have been omitted for brevity. To determine the off-diagonal elements of the Bloch Hamiltonian we define the nearest-neighbor overlap integrals using the well and barrier wavefunctions at the crossing wavevector $k_y = K_y = s \left| K_y \right|$. We can write the intra-cell hopping integral as
\begin{equation}
\label{eq:gamma1}
    \gamma_\text{intra} = \int_{-\infty}^\infty \bra{\psi_b(x)} \hat{H} \ket{\psi_w(x - a)}\text{d} x,
\end{equation}
and the inter-cell hopping integral as
\begin{equation}
\label{eq:gamma2}
    \gamma_\text{inter} = \int_{-\infty}^\infty \bra{\psi_b(x)} \hat{H} \ket{\psi_w(x + L - a)}\text{d} x.
\end{equation}
Combining Eqs.(\ref{eq:SuperlatticeHamiltonian})-(\ref{eq:gamma2}) and performing a nearest-neighbor tight-binding calculation yields the Bloch Hamiltonian in the vicinity of the original band crossing:
\begin{equation}
\label{eq:H}
    \mathcal{H}(\textbf{k}) = \begin{bmatrix}
    s\hbar v_w (k_y - K_y) & f(k_x) \\ f^*(k_x) & s\hbar v_b (k_y - K_y)
    \end{bmatrix},
\end{equation}
where
\begin{equation}
    f(k_x) = \gamma_\text{intra}e^{- i k_x a} + \gamma_\text{inter} e^{i k_x (L-a)}.
\end{equation}
We note that along the superlattice wavevector ($k_x$) axis the Bloch Hamiltonian resembles the Su–Schrieffer–Heeger (SSH) model - the tight-binding model used to describe dimerized atomic chains, e.g., polyacetylene\,\cite{RevModPhys.91.015005}.

As is standard in tight-binding methods the model parameters (i.e., $v_w$, $v_b$, $\gamma_\text{intra}$, and $\gamma_\text{inter}$) can be fit to data, e.g., a numerical calculation of the band structure (see Fig.\,\ref{fig:BipolarToArray}(b)) computed via a transfer matrix (see Appendix\,\ref{sec:transfermatrix} for methods). Whilst the magnitude of the hopping parameters are determined by intra- and inter-cell well and barrier separation, the presence of a band minima at $k_x = 0$ dictates that $\gamma_\text{intra}$ and $\gamma_\text{inter}$ have opposite sign. Furthermore, it can be shown that switching the sign of the wavevector along the guiding potentials ($s$) or the graphene valley index ($s_\text{K}$) flips the sign of the hopping parameters (see Appendix\,\ref{sec:bandgapestimation}). Combining these conditions we can define $\gamma_\text{intra} = s_\text{K} s \gamma_1$ and $\gamma_\text{inter} = s_\text{K} s \gamma_2$ where $\gamma_1 > 0$ and $\gamma_2 < 0$.

\subsection*{Emergence of gapped and tilted Dirac cones}
\label{sec:Emergence}
\begin{figure*}
    \centering
    \includegraphics[width=0.85\textwidth]{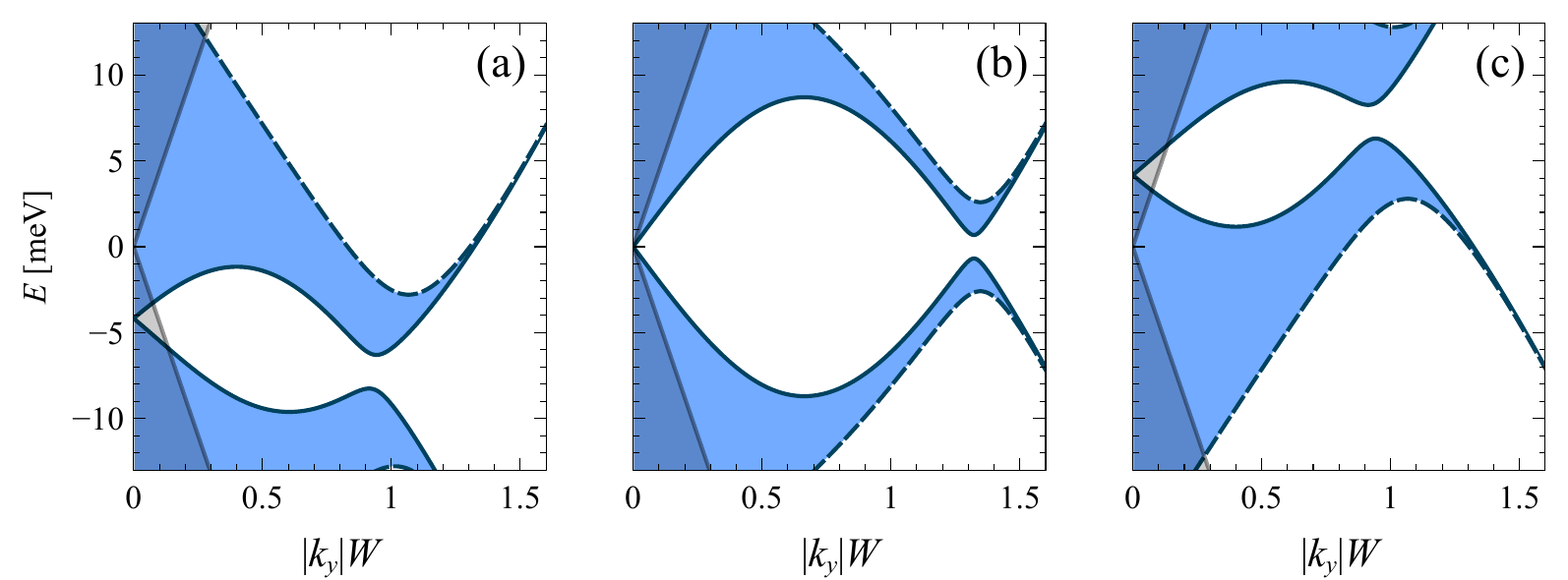}
    \caption{
    Orthographic projections of the band structures of three bipolar arrays as viewed along the superlattice wavevector. In each plot the location of the well within the unit cell is $a = 48\text{nm}$, the well and barrier widths are $W = 15\text{nm}$ and the superlattice unit cell width is $L = 90\text{nm}$. The well and barrier potentials in each panel are: $U_b = 85\text{meV}$ and $U_w = -110\text{meV}$ in panel (a), $U_b = 110\text{meV}$ and $U_w = -110\text{meV}$ in panel (b), and $U_b = 110\text{meV}$ and $U_w = -85\text{meV}$ in panel (c). Varying the well and barrier potentials can be seen to change the tilt of the satellite gapped and tilted Dirac cones within the electronic band structure. The band structures were calculated using a transfer matrix and are plotted in terms of energy $E$ in units of milielectron volts (meV), wavevector along the guiding potentials $k_y$ and superlattice wavevector $k_x$. In the orthographic projection, the band edges are depicted by solid lines ($k_x = 0$) or dashed lines ($k_x = \pm \pi/L$) with intermediate values shaded in blue. The grey areas correspond to energies and wavevectors which support plane wave solutions across the entire potential.
   }
    \label{fig:TiltParam}
\end{figure*}
We now demonstrate the existence of gapped and tilted Dirac cones within the electronic band structure. These Dirac cones can be found at the local band minima, i.e., $k_y = s\left| K_y \right|$ and $k_x = 0$. To capture the quadratic band dispersion of the gapped and tilted Dirac cone, we must expand the Bloch Hamiltonian to the second-order in wavevector. By performing a specific unitary transformation, we can eliminate second-order terms from the Hamiltonian and determine the Dirac cone velocity parameters. We perform the following unitary transformation: $\mathcal{H}'(\textbf{k}) = \mathcal{U}(k_x) \mathcal{H}(\textbf{k}) \mathcal{U}^\dagger(k_x)$ where
\begin{equation}
    \mathcal{U}(k_x) = \frac{1}{\sqrt{2}}\begin{bmatrix}
        e^{i k_x (a - l)} & e^{i k_x l}\\
        i e^{i k_x (a - l)} & -i e^{i k_x l}
    \end{bmatrix},
\end{equation}
with $l = \big[ (\gamma_2 + \sqrt{\left|\gamma_1 \gamma_2\right|})/2(\gamma_1 + \gamma_2)\big] L$. This wavevector-dependent unitary transformation moves our Bloch sums out of the atom gauge (originally defined in Eqs.\,(\ref{eq:BlochSum1}) and (\ref{eq:BlochSum2})). This unitary transformation does not affect the electronic band structure, but can affect the calculation of optical transitions - we discuss this point further in Section\,\ref{sec:THz}. Performing an expansion in terms of $q_x = k_x$ now reveals an effective Bloch Hamiltonian with no second-order wavevector terms:
\begin{equation}
\label{eq:GappedCone}
\mathcal{H}'(\textbf{q}) = s_\text{K} s\frac{E_g}{2} \sigma_z + s \hbar v \big( t q_y \mathbb{I} + q_y \sigma_y + s_\text{K} T q_x \sigma_x \big),
\end{equation}
where $v$ is the modified Fermi velocity, $t$ is a tilt parameter, and $T$ is a Fermi velocity anisotropy factor, $E_g$ is the local band gap and $\textbf{q} = (q_x,q_y)$ is the deviation in wavevector from the Dirac point where $q_y = k_y - K_y$. We note that the offset energy has been omitted for brevity. The effective velocity, tilt, and anisotropy parameters can be expressed through $v = (v_w - v_b)/2$, $t = (v_w +  v_b)/2v$, and $T = \sqrt{\left|\gamma_1 \gamma_2\right|} L/\hbar v$ respectively. Furthermore, the gap parameter can be defined through the nearest-neighbor tight-binding hopping integrals $\left| E_g \right| = 2\left| \gamma_1 + \gamma_2 \right|$. Indeed, Eq.\,(\ref{eq:GappedCone}) is the well-known Dirac cone Hamiltonian possessing a tilted dispersion along the $q_y$ axis, a non-tilted dispersion along the $q_x$ axis, and a local band gap. In contrast to tilted Dirac cone materials formed by crystalline lattices, features such as Dirac cone tilt ($t$) and band gap ($E_g$) can be tuned by varying the applied gate voltages of the bipolar array. Throughout the rest of this paper we will investigate how varying the voltage profiles of the bipolar array results in gate-tunable phenomena stemming from Dirac cone engineering.

It should be noted that previous studies of graphene superlattices have been limited to periodic wells/barriers\,\cite{PhysRevB.77.115446,PhysRevB.81.075438,PhysRevLett.103.046808,10.1140/epjb/e2016-70605-5}, sinusoidal\,\cite{PhysRevLett.103.046809,PhysRevB.79.115427}, periodic even/odd potentials\,\cite{PhysRevLett.101.126804}, or electromagnetic potentials\,\cite{SOMROOB2021114501} which had reflection symmetry, and thus did not open a gap in the Dirac cone. Indeed, we can recover these results by considering the specific case, $a = L/2$, where the bipolar potential possess a reflection plane and the band gap of the tilted cones vanishes ($\gamma_1 = -\gamma_2$ and $E_g = 0$) (see Fig.\,\ref{fig:BipolarToArray}(c)).

Although we have discussed the role of superlattice geometry in opening band gaps in Dirac cones, we emphasize that the full band structure of the bipolar array remains gapless. This is due to gapless Dirac cones that exist at $\textbf{k} = \textbf{0}$ for all superlattice geometries\,\cite{PhysRevLett.101.126804,PhysRevB.77.115446,PhysRevB.81.075438} (see Fig.\,\ref{fig:BipolarToArray}). In this respect, the previously discussed tilted and gapped Dirac cones are satellites to a central gapless Dirac cone. This central Dirac cone is not tilted and has elliptical isoenergy contours which can be fitted by the phenomenological Fermi velocities along the $k_x$ ($v_{c,x} \leq v_\text{F}$), and $k_y$ ($v_{c,y} \leq v_\text{F}$) wavevector axes. The energy offset of the central Dirac cone is equal to the average potential of the bipolar array $W(U_b + U_w)/L$.

\subsection*{Details on the nearest-neighbor tight-binding model}

When applied to crystalline materials, the standard nearest-neighbor tight-binding assumes that each atomic orbital is well-localised to its respective lattice site. In the context of this work, our analytic theory most closely matches the numerical transfer matrix calculations when the individual well and barrier states are sufficiently localised to the confining potential. Outside of the confining well and barrier potentials, the wavefunctions 
corresponding to the crossing wavevector $k_y = K_y$ and crossing energy $E_\text{off}$ are proportional to $e^{\widetilde{\kappa} x}$ (to the left of the potential) or $e^{-\widetilde{\kappa} x}$ (to the right of the potential) where $\widetilde{\kappa} = (1/\hbar v_\text{F})\sqrt{(\hbar v_\text{F} K_y)^2 - E_\text{off}^2}$. Provided that each wavefunction is sufficiently localized within a single superlattice unit cell ($\widetilde{\kappa}  L \gg 1$), we need not consider additional next nearest-neighbor hopping terms. For example, in Fig.\,\ref{fig:BipolarToArray}, where $\left| K_y \right| \! W \approx 1.2$ and $E_\text{off} \approx -10\text{meV}$ it can be checked that $\widetilde{\kappa} L \approx 7$ thereby justifying use of the nearest-neighbour tight-binding model. It should also be noted that the boundary conditions of finite and infinite bipolar arrays are different. Namely, in finite arrays the wavefunction must decay outside of the outermost wells, whereas for the infinite case, the system is subject to the Born-von Karman boundary conditions. Consequently, in finite systems no guided modes exist in the region where $\left| E \right| > \hbar v_\text{F} \! \left| k_y \right|$ (grey regions of Fig.~\ref{fig:BipolarToArray}). Conversely, in the infinite case, guided modes are supported in this region.

\section{Gapped Dirac cones with gate-tunable tilt}
\label{sec:Tilt}
Gapped and tilted Dirac cones have been a topic of intense research. As previously discussed, modifying the degree of tilt leads to drastically different emergent system behavior. As was demonstrated in the context of isolated well and barrier band crossings, the tilt $t$ of Dirac cones in a bipolar array can be modified by tuning the applied gate voltages. For example, as can be seen from Fig.~\ref{fig:TiltParam}, varying the barrier height or well depth tunes the tilt parameter. Interchanging the well depth and barrier height flips the sign of the tilt parameter of the gapped satellite Dirac cones. The experimental ability to continually change the tilt parameter across a broad range of values means that it can be viewed as an additional degree-of-freedom in device applications. As an example of this, in Section\,\ref{sec:LLs}, we explore how varying the tilt of gapped Dirac cones within the electronic band structure will lead to a gate-tunable Landau level spectra.

\begin{figure*}
    \centering
    \includegraphics[width=\textwidth]{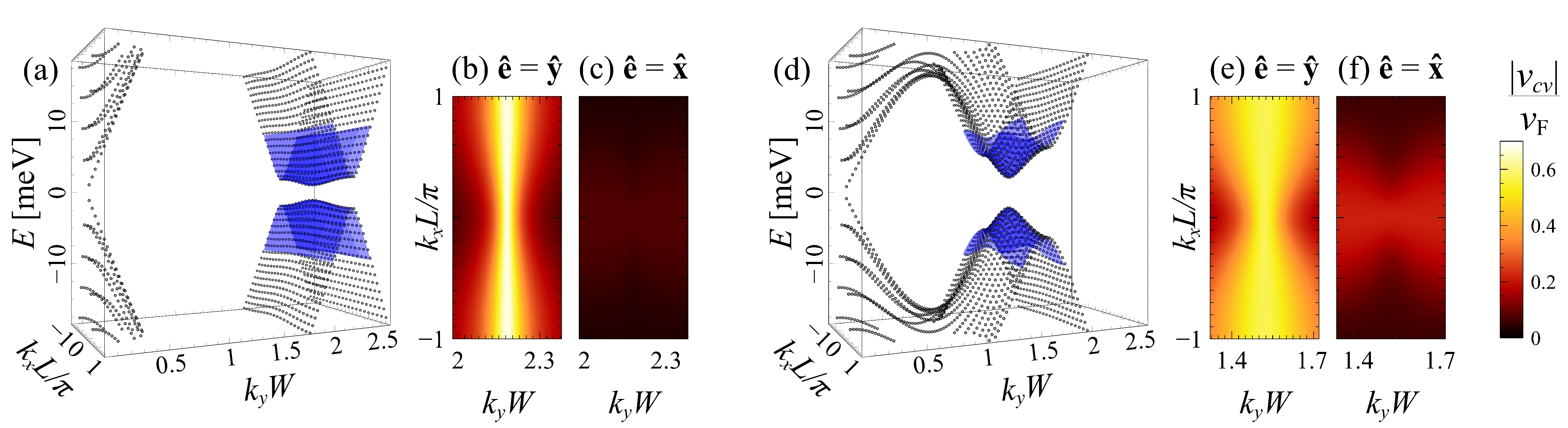}
    \caption{Electronic band dispersions and velocity matrix elements for two bipolar array geometries with well and barrier heights: (a)-(c) $U_b = -U_w = 210\text{meV}$ and (d)-(f) $U_b = -U_w = 175\text{meV}$. In both cases, the superlattice unit cell width is $L = 50\text{nm}$, well and barrier width is $W = 10\text{nm}$ and the separation between the well and barrier within one unit cell is $a = 27.5\text{nm}$. In panels (a) and (d) the full electronic band structure (black dots) is obtained via a transfer matrix calculation and is plotted over a finite range of wavevectors along the guiding potentials ($0 \leq k_y W \leq 2.6$) and the full Brillouin zone along the superlattice axis ($\left| k_x \right| L \leq \pi/L$). In the vicinity of the gapped Dirac cones, we plot the analytic approximation to the full band structure (blue surface) obtained via the superlattice tight-binding model. Using this analytic approximation to the band structure, we plot the absolute value of the velocity matrix element $\left| v_\text{cv}(\textbf{k}) \right|$ for light polarized along ($\hat{\textbf{y}}$ axis) and perpendicular ($\hat{\textbf{x}}$ axis) to the guiding potentials for both cases.}
    \label{fig:VME}
\end{figure*}

The tilted and gapped Dirac cones in Fig.\,\ref{fig:TiltParam} correspond to sub-critically tilted type-I ($\left| t \right| < 1$) gapped Dirac cones. We note that it is possible to increase the tilt parameter further towards critically tilted type-III ($\left| t \right| = 1$) and super-critically tilted type-II ($\left| t \right| > 1$) Dirac cones. We note that for over-tilted Dirac cones (particularly the critically tilted type-III case), one branch of the the electronic band dispersion appears quadratic rather than linear (see Fig.\,\ref{fig:WellBarrierSchem}). When lacking a band gap, these cones are known as three-quarter Dirac points and possess interesting properties such as Landau levels with energy that scales to the four-fifth power of magnetic field strength $B$ and Landau level index $n$, i.e., $E_n \propto (n B)^{4/5}$\,\cite{PhysRevB.96.085430,PhysRevB.99.045409}. The properties of these three-quarter Dirac fermions (with and without a band gap) could be accounted for in our model by adding an effective mass ($m^*$) to either the well or barrier modes. For example, ammending the well dispersion: $s\hbar v_w q_y + \hbar^2 q_y^2/2m^*$ which was originally defined in Eq.\,(\ref{eq:vwell}) adds a quadratic term $(\hbar^2 q_y^2/4m^*)(\mathbb{I} + \sigma_y)$ to the gapped Dirac cone Hamiltonian given in Eq.\,(\ref{eq:GappedCone}). Therefore, in realistic critically (type-III) and super-critically (type-II) tilted Dirac cone materials, the gapped and tilted Dirac cone Hamiltonian may possess an additional quadratic term. This addition to the tilted Dirac cone Hamiltonian goes beyond the standard model used to predict the emergent physics of tilted Dirac cone materials and thus constitutes an interesting avenue of future study.

\section{Tunable Dirac cone gap and terahertz transitions}
\label{sec:THz}
\begin{table}[b!]
\caption{Tight-binding model parameters for a bipolar array characterized by two voltage profiles: $U_b = -U_w = U_0 = 210\text{meV}$ (Model A) and $U_0 = 175\text{meV}$ (Model B). In each case the well and barrier widths are $W = 10\text{nm}$, the superlattice unit cell is $L = 50\text{nm}$ and the well is centered at $a = 27.5\text{nm}$ within the superlattice cell.}
\centering\label{tab:params}
\begin{tabular}{ c || c | c | c | c | c}
 Model & $U_0$/meV & $\left| K_y \right| W$ & $v_0/v_\text{F}$ & $\gamma_1$/meV & $\gamma_2$/meV\\
 \hline \hline
  A & 210 & 2.18 & 0.68 & 0.52 & -1.56\\
  \hline 
  B & 175 & 1.52 & 0.57 & 1.86 & -3.99
\end{tabular}
\end{table}
In traditional tight-binding models, the atomic orbital wavefunctions are not known, as a result, model parameters such as hopping integrals are fit to experiment. For the case of equal well and barrier strengths ($U_b = -U_w = U_0$) the band crossing occurs at zero-energy, resulting in non-tilted ($v_w = -v_b = v_0$ and $t = 0$) Dirac cones in the electronic band structure. In this case, the well and barrier wavefunctions can be found analytically. These wavefunctions yield a transcendental equation for the crossing wavevector $K_y$, analytic expressions for the well and barrier group velocities $v_0$ as well as the hopping parameters $\gamma_1$ and $\gamma_2$ (see Appendices\,\ref{sec:ZeroEnergyStates} and \,\ref{sec:bandgapestimation}). For example, let us consider a bipolar array characterized by the geometry parameters $W = 10\text{nm}$, $L = 50\text{nm}$, and $a = 27.5\text{nm}$. We consider realistic potential strengths~\cite{PhysRevLett.123.216804}, e.g., $U_0 = 210\text{meV}$ and $U_0 = 175\text{meV}$ in models A and B respectively. For these two models, we can derive values for the tight-binding parameters (see Table\,\ref{tab:params}). Substituting these parameters into the effective Bloch Hamiltonian (see Eq.\,(\ref{eq:H})) provides an accurate match to the electronic band structure obtained via a transfer matrix (see Fig.\,\ref{fig:VME}(a) and (d)).

Using expressions for the tight-binding hopping parameters, we can derive an expression that directly determines the Dirac cone band gap from the bipolar array geometry:
\begin{equation}
\label{eq:Gap}
    \left| E_g \right| = E_0 \sinh \bigg( \frac{\left| K_y \right| \! \left| L - 2a \right| }{2} \bigg) e^{-\left| K_y  \right| L /2},
\end{equation}
where $E_0 = 4\hbar^3 v^3_\text{F}  \!\left| K_y \right|\! \widetilde{K}^2 e^{\left| K_y \right| W}/ U_0^2 (1 + \left| K_y \right|\!W )$, and $\left|K_y \right|$ satisfies the transcendental equation 
$\widetilde{K} = -\left| K_y \right| \tan{(\widetilde{K} W)}$
with $\widetilde{K} = (1/\hbar v_\text{F})\sqrt{U_0^2 - (\hbar v_\text{F} K_y)^2}$ (see Appendix\,\ref{sec:bandgapestimation}). By varying the voltage profile of a bipolar array we can tune the Dirac cone band gap within the THz regime: $E_g = 0.50\text{THz}$ for model A and $E_g = 1.03\text{THz}$ for model B.

The possibility of tuning the local band gap of the Dirac cones into the THz regime provides a route to THz applications arising from interband transitions. We assume that the offset gate voltage of the superlattice places the Fermi level within the band gap of the gapped Dirac cones. Upon illumination of light, a photon of energy $h\nu$ can excite an electron from the valence band up to an empty state in the conduction band provided that the photon energy is equal to the energy separation of the states $h \nu = E_+(\textbf{k}) - E_-(\textbf{k})$. The probability of optical transitions between some state at some wavevector $\textbf{k}$ is determined by the absolute value square of the velocity matrix element (VME) $\left| v_\text{cv}(\textbf{k}) \right|^2$ where
\begin{equation}
    v_\text{cv}(\textbf{k}) = \bra{\Psi_\mp(\textbf{k})} \hat{\textbf{e}} \cdot \textbf{v}(\textbf{k}) \ket{\Psi_\pm(\textbf{k})}.
\end{equation}
Here, $\ket{\Psi_\pm(\textbf{k})}$ are the conduction ($+$) and valence ($-$) states of the low-energy Bloch Hamiltonian given in Eq.\,(\ref{eq:H}), $\textbf{v}(\textbf{k})$ is the velocity operator, and $\hat{\textbf{e}} = e_x \hat{\textbf{x}} + e_y \hat{\textbf{y}}$ is the polarization vector of light. We note that we utilize the eigenstates $\ket{\Psi_\pm(\textbf{k})}$ and velocity operator $\textbf{v}(\textbf{k})$ are defined in the basis of well and barrier Bloch sums $\ket{\Phi_w}$ and $\ket{\Phi_b}$. Considering that this Bloch Hamiltonian is in the so-called atom gauge, the velocity operator can be conveniently determined through the gradient approximation $\textbf{v}(\textbf{k}) = (1/\hbar)\boldsymbol{\nabla}_\textbf{k}\mathcal{H}(\textbf{k})$\,\cite{10.21468/SciPostPhysCore.6.1.002}.

In Fig.\,\ref{fig:VME}, we plot the absolute value of the VME for a range of wavevectors in the vicinity of the gapped Dirac cones. Here, we consider a single bipolar array with two different voltage profiles, i.e., models A and B with parameters given in Table\,\ref{tab:params}. Optical transitions are supported in the vicinity of the gapped Dirac cones for all polarizations of light. For light polarized along the guiding potentials ($\hat{\textbf{e}} = \hat{\textbf{y}}$) the max value of the VME is $v_0$ (for $k_y = K_y$) whilst the light polarized along the array axis ($\hat{\textbf{e}} = \hat{\textbf{x}}$) the max value of the VME is $\left| (L-a)\gamma_2 - a \gamma_1\right|/\hbar$ (for $k_x = 0$). For light polarized along the guiding potentials ($\hat{\textbf{e}} = \hat{\textbf{y}}$) we see that optical transitions are guaranteed for photons with energies spanning $2 \! \left| \gamma_1 + \gamma_2 \right|$ to $2 \! \left| \gamma_1 - \gamma_2 \right|$. Varying the voltage profile of the bipolar array allows for convenient control over this bandwidth after device fabrication. In this frequency regime, there appears to be a preference to absorb photons polarized along the $\hat{\textbf{y}}$ axis, thus, a bipolar array in graphene could be used as a component in a tunable thin-film THz polarizer.

In Fig.\,\ref{fig:VME} we clearly observe the optical momentum alignment phenomenon in which photoexcited electrons are aligned with wavevector perpendicular to the plane of polarizing light. Combining this momentum alignment phenomenon with the tilt\,\cite{wild2023optical} or warping\,\cite{Saroka2022} of the satellite Dirac cones could result in the spatial separation of photoexicted carriers belonging to different satellite cones (differentiated by the index $s$). The optical properties of gapless and gapped tilted Dirac cones are discussed in detail within Refs.\,\cite{wild2021optical, PhysRevB.106.165404} and \cite{PhysRevB.103.165415} respectively. 

We can also investigate the absorption of right-handed $\hat{\textbf{e}}_\circlearrowright = \big( \hat{\textbf{x}} + i \hat{\textbf{y}}\big)/\sqrt{2}$ and left-handed $\hat{\textbf{e}}_\circlearrowleft = \big( \hat{\textbf{x}} - i \hat{\textbf{y}}\big)/\sqrt{2}$ circularly polarized light. For demonstrative purposes, we evaluate the absolute value of the VME for right-handed circularly polarized light at the apex of the gapped Dirac cones $\textbf{k} = (0,K_y)$ obtaining: $\left| a\gamma_1 - (L-a)\gamma_2 + s\hbar v_0\right|/\sqrt{2} \hbar$. In this case, illumination from right-handed polarized light will generate more photoexcited carriers in satellite Dirac cones with index $s = 1$. If the well depth and barrier heights are not equal, these gapped Dirac cones will be tilted in a direction dictated by the sign of $s$ (see Fig.\,\ref{fig:BipolarToArray}(c)). The group velocities resulting from the tilted band structures will result in a photocurrent along the waveguide axis. The direction of the photocurrent will be determined by the handedness of the circularly polarized light. This phenomena is somewhat similar to the ratchet photocurrent predicted for graphene supperlattices formed by periodic
  strain\,\cite{PhysRevB.84.235440}.

It is noted that whilst the gapped satellite Dirac cones do not support the absorption of photons with energy less than the band gap ($2 \left| \gamma_1 + \gamma_2 \right|$), the central gapless Dirac cone will support the absorption of photons with arbitrarily low photon energies.

\section{Designer Landau level spectra}
\label{sec:LLs}
In this section, we consider a typical bipolar array geometry with an electronic band structure containing central gapless Dirac cones and gapped satellite tilted Dirac cones. We assume that the voltage profile of the superlattice has been selected so that the satellite cones are sub-critically tilted (type-I, $\left| t \right| < 1$), see Fig.\,\ref{fig:TiltParam}. In the presence of an external magnetic field oriented normal graphene sheet (with field strength $B$), the energy levels of the charge carriers become quantized in to Landau levels (LLs). For the gapless central Dirac cones, it is well-known that the LLs take on the energy spectra 
\begin{equation}
\label{eq:gaplessLLs}
    E_{n_c} = \text{sign}(n_c) \sqrt{2 \hbar \bar{v}_c^2 e B \! \left| n_c \right|},
\end{equation}
where $\bar{v}_c = \sqrt{v_{c,x}v_{c,y}}$ is the effective Fermi velocity of the central Dirac cone, $n_c$ is a LL index and $e$ is the elementary charge. Each LL has 4-fold degeneracy arising from each graphene valley ($s_\text{K} = \pm 1$) and electron spin. 

\begin{figure}[t]
    \centering
    \includegraphics[width=0.48\textwidth]{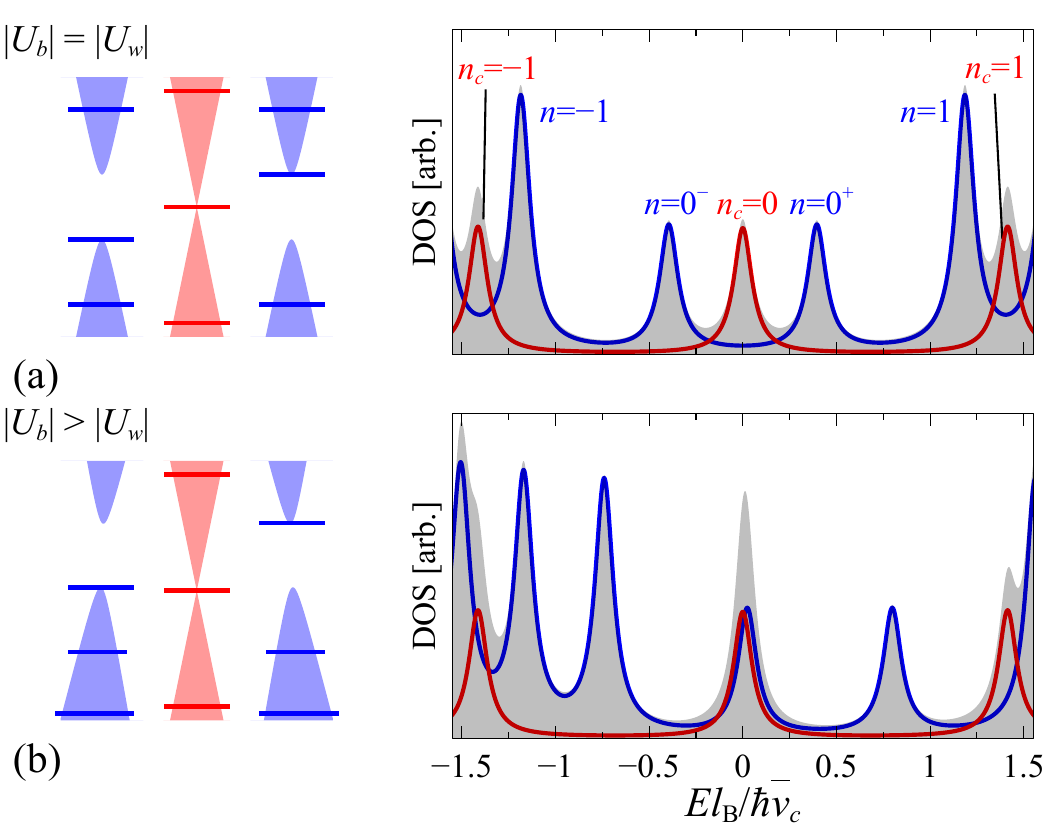}
    \caption{Schematic of the Landau level spectra of a bipolar array in graphene under an external magnetic field for equal well and barrier heights ($\left| U_b \right| = \left| U_w \right|$) in panel (a) and unequal well and barrier heights in panel ($\left| U_b \right| > \left| U_w \right|$) in panel (b). The total density of states has been sketched in grey which is the sum of the contributions from the massive (blue, Landau level index $n$) and massless (red, Landau level index $n_c$) Dirac cones. The energy axis is normalized according to the effective Fermi velocity of the central massless Dirac cone ($\bar{v}_c$) so that the $n_c = -1, 0$ and $1$ Landau level energies take on the values $-\sqrt{2}\hbar \bar{v}_c/l_B, 0$ and $\sqrt{2} \hbar \bar{v}_c/l_B$ where $l_B = \sqrt{\hbar/eB}$ is the magnetic length. In panel (a) the satellite Dirac cones are non-tilted $t = 0$ and have the same offset energy as the central cone whilst in panel (b) the satellite cones are tilted and are offset from the central cone causing overlap of LLs. This figure has been plotted for arbitrary field strength and Dirac cone parameters and each Landau level has been modeled as a Lorentzian with a finite width.}
    \label{fig:LLs}
\end{figure}

Assuming that the applied gate voltages are selected such that the gapped satellite Dirac cones are of type-I, the LL spectra of the tilted gapped satellite Dirac cones, described by the Hamiltonian given in Eq.\,(\ref{eq:GappedCone}), take the form 
\begin{equation}
\label{eq:LLGapped}
    E_{n} = \text{sign}(n)\sqrt{2 \hbar \bar{v}^2 e \lambda^3 B \left|n\right| + \bigg(\frac{\lambda E_g}{2}\bigg)^2},
\end{equation}
for LL index $\left| n \right| \geq 1$\,\cite{PhysRevB.91.085112}, where for the bipolar array $\bar{v} = v \sqrt{T}$ and $\lambda = \sqrt{1-t^2}$ with each LL having 8-fold degeneracy from each graphene valley ($s_\text{K} = \pm 1$), satellite ($s = \pm 1$) and spin. It should be noted that in the presence of a gap, the zeroth LL splits into sub-levels at the band edges $E_{0^+} = \lambda E_g/2$ (when $s = 1$) or $E_{0^-} = -\lambda E_g/2$ (when $s = -1)$ such that the degeneracy of each sub-level is half the other LLs (see Appendix\,\ref{sec:LLstates}). By modifying the applied voltages of the electrostatic superlattice, the tilt ($t$) and band gap ($E_g$) of the gapped satellite Dirac cones can be tuned. We note that the offset energy of the satellite Dirac cones is different to the offset energy of the central Dirac cone. Varying the applied gate voltages of the bipolar array tunes the offset energies between the gapless and gapped LL spectra; this is illustrated schematically in Fig.\,\ref{fig:LLs}. These theoretical results are consistent with a previous numerical study into the formation of Landau levels in graphene superlattices (see Ref.\,\cite{PhysRevLett.103.046808}). In turn, this allows designer LL spectra via Dirac cone engineering which would be measurable in magneto-resistance experiments or through magneto-optic transitions.

\section{Conclusion}
Research in to the physics of gapless and gapped tilted Dirac cone materials\,\cite{doi:10.1143/JPSJ.79.114715,PhysRevB.96.155418,PhysRevB.103.165415,PhysRevB.103.125425,wild2021optical,C7CP03736H,PhysRevB.97.235113,PhysRevB.97.235440,Sengupta_2018,PhysRevB.99.235413,PhysRevB.102.045417,doi:10.1143/JPSJ.77.064718,PhysRevB.91.195413,PhysRevB.96.235405,PhysRevB.98.195415,PhysRevB.99.035415,ng2021mapping} is in its infancy, having been inspired by the prediction tilted Dirac cones in 8-$Pmmn$ borophene, a boron monolayer. In each of these works the tilt parameter takes on a fixed value that is assumed to be predetermined by rigid lattice geometries. In this work we propose a feasible method to engineer gapped and tilted Dirac cones in a lateral graphene superlattice. In stark contrast to crystalline atomic monolayers, the electronic band structure of a graphene superlattice can be modified by varying the applied voltage profile - this provides a practical means to control the tilt parameter and band gap of Dirac cones. 

Whilst this work has been focused on the study of one-dimensional lateral superlattices in graphene, we note that two-dimensional graphene superlattices\,\cite{PhysRevLett.101.126804} may also provide a viable platform to realize designer gapped and tilted Dirac cones. It is also noted that although in this section we have considered lateral superlattices applied to graphene, it may also be possible to consider other superlattice geometries made possible through strain\,\cite{PhysRevB.84.235440,PhysRevB.82.155417}, doping\,\cite{Wang2023}, or electromagnetic fields\,\cite{SOMROOB2021114501,condmat8010028}. Furthermore, we need not limit our substrate to graphene, superlattices could also be considered for other two-dimensional systems such as bilayer graphene\,\cite{doi:10.1098/rsta.2010.0218}, silicene\,\cite{PhysRevB.90.125444} or eventually, two-dimensional materials that already host tilted Dirac cones in the electronic band structure, i.e., 8-$Pmmn$ borophene\,\cite{Xu_2023}.    

The tilted and gapped Dirac cones within a lateral graphene superlattice can be engineered to give desirable device characteristics - as examples of this we discussed tunable THz applications and designer Landau level spectra. It was shown that the a lateral graphene superlattice can be engineered to absorb THz photons within a narrow bandwidth. This bandwidth can be tuned post-fabrication by varying the voltage profile of the superlattice. We hope that this work will encourage the use of lateral graphene bipolar superlattices in the design of novel THz devices.

\section*{Acknowledgments}
This work was supported by the EU H2020-MSCA-RISE projects TERASSE (Project No. 823878) and CHARTIST (Project No. 101007896). A.W. was supported by a UK EPSRC PhD studentship (Ref. 2239575) and by the NATO Science for Peace and Security project NATO.SPS.MYP.G5860. 
E.M. acknowledges financial support from the Royal Society International Exchanges grant number IEC/R2/192166. M. E. P. acknowledges support from UK EPSRC (Grant No. EP/Y021339/1).  

\appendix

\section{Zero-energy states of quantum wells and barriers in graphene}

\label{sec:ZeroEnergyStates}
In this appendix, we present expressions for zero-energy states for quantum wells and barrier in graphene. Owing to the symmetry of the confining potential $U(x) = U(-x)$, it is convenient to re-write Eq.\,(\ref{eq:DiracEffective}) in the symmeterized basis, i.e., $\ket{\psi(x)} = \big[ \psi_1(x), \psi_2(x)\big]^\text{T}$ where $\psi_1(x) = [\psi_A(x) + i\psi_B(x)]/\sqrt{2}$ and $\psi_2(x) = [\psi_A(x) - i\psi_B(x)]/\sqrt{2}$. The spinor components satisfy the following simultaneous equations:
\begin{equation}
\label{eq:Sim1}
    \bigg[\frac{U(x) - E}{\hbar v_\text{F}} - s_\text{K} k_y\bigg] \psi_1(x) + \partial_x \psi_2(x) = 0,   
\end{equation}
and
\begin{equation}
\label{eq:Sim2}
    -\partial_x \psi_1(x) + \bigg[\frac{U(x) - E}{\hbar v_\text{F}} + s_\text{K} k_y\bigg] \psi_2(x) = 0.
\end{equation}
In conjunction with Eq.\,(\ref{eq:potentials}), we define three regions of the square potential: I ($x < -W/2$), II ($-W/2 \leq x \leq W/2$) and III ($x > W/2$). The total wavefunction is obtained by solving Eqs.\,(\ref{eq:Sim1}) and (\ref{eq:Sim2}) in each region of the potential and matching the spinor components at the boundaries. We note that for the case of graphene, in contrast to traditional free-electron quantum well problems, it is not necessary to match the derivative of the spinor components.

For a quantum barrier with height $\pi/2 < U_{0} W/\hbar v_\text{F} < 3\pi/2$ there are two zero-energy solutions in each graphene valley ($s_\text{K} = \pm 1$) - one has a positive wavevector and the other has a negative wavevector $k_y = s \left| K_y \right|$ with $s = \pm 1$. It can be seen from Eqs.\,(\ref{eq:Sim1}) and (\ref{eq:Sim2}) that interchanging the graphene valley index is mathematically equivalent to changing the sign of the wavevector along the guiding potential. We first solve for a quantum barrier in graphene, for the case $s_\text{K}s = 1$ where the zero-energy wavefunction takes the form
\begin{equation}
\label{eq:zero-energystateeq1}
     \ket{\psi_b^\text{I}(x)} = \frac{1}{N} e^{\left| K_y \right| W/2} \sin(\widetilde{K} W/2) \begin{pmatrix}
     1 \\ 1 
     \end{pmatrix} e^{\left| K_y \right| x},
 \end{equation}
\begin{equation}
 \ket{\psi_b^\text{II}(x)} = \frac{1}{N} \begin{bmatrix}
     \tan(\widetilde{K} W/2)\cos(\widetilde{K} x)\\
     -\sin(\widetilde{K} x)
 \end{bmatrix},
 \end{equation}
 and
 \begin{equation}
 \label{eq:zero-energystateeq2}
     \ket{\psi_b^\text{III}(x)} = \frac{1}{N} e^{\left| K_y \right| W/2} \sin(\widetilde{K} W/2) \begin{pmatrix}
     1 \\ -1 
     \end{pmatrix} e^{-\left| K_y \right| x},
 \end{equation}
where the effective wavevector in the barrier is
\begin{equation}
    \widetilde{K} = \frac{1}{\hbar v_\text{F}} \sqrt{U_0^2 - (\hbar v_\text{F} K_y)^2},
\end{equation}
and the normalization factor is
\begin{equation}
    N = \sqrt{\frac{(\left| K_y \right| W + 1)}{2\left| K_y \right|}}\sec(\widetilde{K} W/2).
\end{equation}
The zero-energy states occur at wavevector $s \left| K_y \right|$ where $\left| K_y \right|$ is the solution to the transcental equation
 \begin{equation}
 \label{eq:Transcendental}
     \widetilde{K} = - \left| K_y \right| \tan(\widetilde{K} W).
 \end{equation}
 Here, we have centered the barrier at the co-ordinate $x = 0$, however, the barrier can be offset by setting $x \to x - x_0$. We can see from Eqs.\,(\ref{eq:Sim1}) and (\ref{eq:Sim2}) that a wavefunction for the case $s_\text{K}s = -1$ can be obtained from the $s_\text{K}s = 1$ wavefunction through the operation $\sigma_x \!\ket{\psi_{w/b}(-x)}$. It can also be seen from Eqs.\,(\ref{eq:Sim1}) and (\ref{eq:Sim2}) that the zero-energy ($E = 0$) wavefunction for a well can be related to that of a barrier $\ket{\psi_w(x)} = \sigma_x \! \ket{\psi_{b}(x)}$ provided the well depth is equal to the barrier height $U_b = -U_w = U_0$.

We can obtain the group velocity of the well and barrier dispersions at the crossing point by calculating expectation of the velocity operator $v_{w/b} = \int \bra{\psi_{w/b}(x)} \hat{v} \ket{\psi_{w/b}(x)}\text{d}x$ where in the symmeterized basis the velocity operator is defined as $\hat{v} = -s_\text{K} v_\text{F} \sigma_z$ where $\sigma_z$ is the third Pauli matrix. Performing this calculation yields the barrier and well group velocities $v_w = -v_b = v_0$ where
\begin{equation}
    v_0 = \frac{s\hbar v_\text{F}^2 \left| K_y \right|}{U_0}.
\end{equation}

\section{Transfer matrix method for the bipolar array in graphene}
\label{sec:transfermatrix}
To support the theoretical predictions of our work we provide a transfer matrix model that can be used to numerically calculate the electronic band structure of the bipolar array in graphene. The employed transfer matrix model is based on earlier works used to derive the electronic band structure of simpler graphene superlattices\,\cite{PhysRevB.77.115446,PhysRevB.81.075438}. The general theory of the transfer matrix method for Dirac systems is discussed in Ref.\,\cite{PhysRevC.35.2262}.

The bipolar array has a superlattice unit cell consisting of four regions ($n = 1$ to $4$) with potential ($U_n$) between the co-ordinates $x_{n-1} \leq x \leq x_n$. For consistancy with the theoretical model the potentials take the values $U_1 = U_b$, $U_3 = U_w$ and $U_2 = U_4 = 0$ whilst the boundaries take the values $x_0 = -W/2$, $x_1 = W/2$, $x_2 = a - W/2$, $x_3 = a + W/2$ and $x_4 = L - W/2$. The wavefunction in region $n$ in unit cell $j$ can be found by solving Eq.\,(\ref{eq:DiracEffective}) for a constant potential $U_n$ yielding $\ket{\psi_n(x)} = \boldsymbol{\Omega}_n(x)\big( \alpha^{(j)}_{n}, \beta^{(j)}_{n}\big)^\text{T}$ where $\alpha^{(j)}_{n}$ and $\beta^{(j)}_{n}$ are wavefunction components and T is the transpose operator. For the case $\left| U_n - E \right| \geq \hbar v_\text{F} \! \left| k_y \right|$ we obtain guided mode solutions encoded by the matrix
\begin{equation}
\label{eq:Transfer1}
    \boldsymbol{\Omega}_n(x) = \begin{pmatrix}
        e^{i \widetilde{k}_n x} & e^{-i \widetilde{k}_n x}\\
        \Lambda_{n,+} e^{i \widetilde{k}_n x} & \Lambda_{n,-} e^{-i \widetilde{k}_n x}
    \end{pmatrix},
\end{equation}
where the effective wavevector is
\begin{equation}
    \widetilde{k}_n = \frac{1}{\hbar v_\text{F}} \sqrt{(U_n - E)^2 - (\hbar v_\text{F} k_y)^2},
\end{equation}
and
\begin{equation}
\label{eq:Transfer3}
    \Lambda_{n,\pm} = \frac{E - U_n}{\hbar v_\text{F}\big(\!\pm \!\widetilde{k}_n - i k_y\big)},
\end{equation}
which are defined for a single graphene valley ($s_\text{K} = 1$) up to a normalization factor. We note that for the case $\left| U_n - E \right| < \hbar v_\text{F} \! \left| k_y \right|$ the wavefunction decays - this is achieved by replacing $\widetilde{k}_n$ with $i \widetilde{\kappa}_n$ in Eqs.(\ref{eq:Transfer1})-(\ref{eq:Transfer3}) where
\begin{equation}
    \widetilde{\kappa}_n = \frac{1}{\hbar v_\text{F}} \sqrt{ (\hbar v_\text{F} k_y)^2 - (U_n - E)^2}.
\end{equation}

To obtain the total wavefunction of the bipolar array we sequentially satisfy each boundary condition in the superlattice potential. By matching all boundary conditions within a single unit cell, we can relate the wavefunction in unit cell $j$ to the wavefunction in the neighboring unit cell $j+1$
\begin{equation}
     \begin{pmatrix}
        \alpha^{(j)}_1\\
        \beta^{(j)}_1
    \end{pmatrix} = \textbf{T} \begin{pmatrix}
        \alpha^{(j+1)}_1\\
        \beta^{(j+1)}_1
    \end{pmatrix},
\end{equation}
where the transfer matrix is defined as
\begin{equation}
    \textbf{T} = \Bigg[ \prod_{n = 1}^3 \boldsymbol{\Omega}_n^{-1}(x_{n}) \boldsymbol{\Omega}_{n+1}(x_{n})\Bigg] \boldsymbol{\Omega}_4^{-1}(x_{4}) \boldsymbol{\Omega}_{1}(x_{0}).
\end{equation}
In conjunction with the theoretical model (see Eq.\,(\ref{eq:SuperlatticeHamiltonian})) this superlattice unit cell is repeated $N$ times leading to the expression
\begin{equation}
    \begin{pmatrix}
        \alpha^{(j)}_1\\
        \beta^{(j)}_1
    \end{pmatrix} = \textbf{T}^N \begin{pmatrix}
        \alpha^{(j+N)}_1\\
        \beta^{(j+N)}_1
    \end{pmatrix}.
\end{equation}
Due to the translational invariance of the superlattice the spinor components $\alpha_1^{(j)}$ and $\beta_1^{(j)}$ are equivalent to $\alpha_1^{(j+N)}$ and $\beta_1^{(j+N)}$ respectively. This places a constraint on the transfer matrix ($\textbf{T}^N = 1$) yielding the eigenvalues $e^{\pm 2 i j \pi/N}$. For the case of an infinite superlattice ($N \to \infty$) we label the continuum of eigenvalues $e^{\pm i k_x L}$ with the superlattice wavevector ($k_x = 2 \pi j /NL$). As discussed in Ref.\,\cite{PhysRevC.35.2262} The electronic band structure is found by searching for energy ($E$) and wavevector ($\textbf{k} = \big( k_x, k_y\big)$) values that satisfy the condition
\begin{equation}
    2 \cos(k_x L) = \text{Tr}\big( \textbf{T}\big).
\end{equation}

 \section{Tight-binding hopping parameters and estimation of Dirac cone band gap}
\label{sec:bandgapestimation}
For the case of the bipolar array that lacks a reflection plane in the superlattice, gapped Dirac cones appear. The band gap of these Dirac cones is given by the magnitude twice the sum of the intra- and inter-cell hopping integrals. For the case of equal well depth and barrier height ($U_b = -U_w = U_0$) the well and barrier dispersions cross at zero-energy. In this case, we can obtain analytic expressions for $\gamma_\text{intra}$ and $\gamma_\text{inter}$ by utilizing the analytic zero-energy solutions to the square well and barrier in graphene provided in Appendix\,\ref{sec:ZeroEnergyStates}. 

We will begin by calculating the intra-cell hopping parameter for the specific case of $s_\text{K}s = 1$. Without loss of generality we specify the unit cell ($j = 0$) and subtitute the superlattice Hamiltonian $\hat{H}$ (see Eq.\,(\ref{eq:SuperlatticeHamiltonian})) into Eq.\,(\ref{eq:gamma1}). Removing all negligible terms yields
\begin{equation}
    \begin{split}
        \gamma_\text{intra} = & \int_{-\infty}^\infty \bra{\psi_b(x )} U_b(x ) \ket{\psi_w(x - a)}\text{d} x\\ &+ \int_{-\infty}^\infty \bra{\psi_b(x)} \hat{H}_\text{G} + U_w(x- a) \ket{\psi_w(x - a)}\text{d} x.
    \end{split}
\end{equation}
We note that the term $[\hat{H}_\text{G} + U_w(x - a)]\ket{\psi_w(x - a)}$ is the eigenvalue problem given in Eq.\,(\ref{eq:DiracEffective}). As the wavefunction $\ket{\psi_w(x-a)}$ corresponds to a zero-energy state this term vanishes. Inputting the definition for a quantum barrier (defined through Eq.\,(\ref{eq:potentials})) yields the simplified expression for the hopping parameter
\begin{equation}
\label{eq:gamma1b}
    \gamma_\text{intra} = U_0 \int_{-W/2}^{W/2} \bra{\psi_b(x )} \ket{\psi_w(x - a)}\text{d} x.
\end{equation}
Inputting the analytic solutions for the zero-energy states (see Eqs.\,(\ref{eq:zero-energystateeq1})-(\ref{eq:Transcendental}) and $\ket{\psi_w(x)} = \sigma_x\ket{\psi_{b}(x)}$) into Eq.\,(\ref{eq:gamma1b}) and solving the resultant integral yields an expression for the intra-cell hopping parameter $\gamma_\text{intra} = \gamma_1$ (when $s_\text{K}s=1$) where
\begin{equation}
    \gamma_1 = \frac{\hbar^3 v_\text{F}^3 \left| K_y \right| \widetilde{K}^2}{U_0^2 (1 + \left| K_y \right| W)} e^{\left| K_y \right|(W-a)}.
\end{equation}
Carrying out the same procedure for the inter-cell hopping parameter yields $\gamma_\text{inter} = \gamma_2$ (when $s_\text{K}s=1$) where $\gamma_2 = -\gamma_1 \exp[\left| K_y \right|(2a-L)]$. For the alternate case $s_\text{K}s = -1$, the well and barrier wavefunctions are modified as $\sigma_x \ket{\psi_{w/b}(-x)}$ (see Appendix\,\ref{sec:ZeroEnergyStates}). Substituting this transformation into Eqs.\,(\ref{eq:gamma1}) and (\ref{eq:gamma2}) reveals that flipping the sign of $s_\text{K}s$ is mathematically equivalent to switching the well and barrier positions. Thus, for the symmetric case ($a = L/2$), switching the sign of $s_\text{K}s$ simply interchanges the inter- and intra-cell hopping parameters. In general, when $a \neq L/2$, we obtain $\gamma_\text{intra} = s_\text{K}s \gamma_1$ and $\gamma_\text{inter} = s_\text{K} s \gamma_2$. We can then obtain the local band gap of the gapped Dirac cones thorugh $\left| E_\text{g} \right| = 2\left| \gamma_1 + \gamma_2 \right|$ which yields the solution given in Eq.\,(\ref{eq:Gap}) of the main text.

\section{Landau level wavefunctions in massive tilted Dirac cones}
\label{sec:LLstates}
In the presence of a magnetic field, we substitute the vector potential into the Hamiltonian given in Eq.\,(\ref{eq:GappedCone}) using the identity $\hat{\textbf{q}} \to \hat{\textbf{q}} + e\textbf{A}/\hbar$. Here, the wavevector operators take on the value $\hat{q}_{x,y} = -i\partial_{x,y}$ whilst the vector potential $\textbf{A} = -Bx\hat{\textbf{y}}$ describes a magnetic field normally incident on the system. In this gauge, the Hamiltonian is solved by the wavefunction $\ket{\Psi_n(x,y)} = e^{i q_y y}\ket{\Psi_n(x)}$ resulting in the eigenvalue problem $\hat{\mathcal{H}}_B \ket{\Psi_n(x)} = E_n \ket{\Psi_n(x)}$ with
\begin{equation}
\begin{split}
\label{eq:Hb}
   \hat{\mathcal{H}}_B =\: & s_\text{K} s \frac{E_g}{2}\sigma_z + s v ( \hbar q_y - eBx )(t \mathbb{I} + \sigma_y) \\&- i s_\text{K} s \hbar v T \frac{\partial}{\partial x} \sigma_x, 
\end{split}
\end{equation}
where $n$ is the LL index, $E_n$ is the Landau level energy and $\ket{\Psi_n(x)}$ is the associated LL wavefunction. 

Whilst this problem can be solved using a generalized chiral operator\,\cite{PhysRevB.91.085112}, we solve it using an approach previously outlined for gapless tilted Dirac cones in Ref.\,\cite{doi:10.1143/JPSJ.79.044708} which we have adapted for the gapped case. For LLs with index $\left| n \right| \geq 1$ the energy spectra is defined in Eq.\,(\ref{eq:LLGapped}) whilst the wavefunctions take the form
\begin{equation}
\label{eq:LL1}
\begin{split}
    \ket{\Psi_n(x)} = &\frac{e^{-X_n^2/2}}{N_n}\Bigg[ (2s \varepsilon_n + \lambda \varepsilon_g)\begin{pmatrix}
        1 + \lambda \\ -it
    \end{pmatrix} h_{\left| n \right|} (X_n)\\& -2i \sqrt{2\lambda^3\left| n \right|}\begin{pmatrix}
        it \\ 1 + \lambda
    \end{pmatrix} h_{\left| n \right| - 1} (X_n) \Bigg],
\end{split}
\end{equation}
for $s_\text{K} = 1$. In these expressions, for brevity, we have utilized dimensionless variables for the energy spectra $\varepsilon_n = E_n l_B/\hbar \bar{v}$ and band gap $\varepsilon_g = E_g l_B/\hbar \bar{v}$ which are defined through the magnetic length $l_B = \sqrt{\hbar/eB}$. In addition, we have utilized the scaled and translated co-ordinate
\begin{equation}
\label{eq:LL1b}
   X_n = \sqrt{\lambda} \bigg( \frac{x}{l_B \sqrt{T}} - \frac{q_y l_B}{\sqrt{T}} - \frac{s \varepsilon_n t}{\lambda^2} \bigg),
\end{equation}
the normalization factor
\begin{equation}
    N_n = \sqrt{2(1 + \lambda)}\sqrt{(2s\varepsilon_n + \lambda \varepsilon_g)^2 + 8\lambda^3\left| n \right| },
\end{equation}
and the normalized Hermite polynomials
\begin{equation}
    h_m(X_n) = \bigg( \frac{\lambda}{\pi T} \bigg)^{\frac{1}{4}}\frac{1}{\sqrt{2^m l_B m!}} H_m(X_n),
\end{equation}
where $H_m()$ are Hermite polynomials. For the second graphene valley ($s_\text{K} = -1$) the LL wavefunction takes the form $-i \sigma_y \! \ket{\Psi_n}$. As discussed in the main text each LL with index $\left| n \right| \geq 1$ has 8-fold degeneracy arising from spin, graphene valley ($s_\text{K} = \pm 1$) and satellite Dirac cones ($s = \pm 1$).

As discussed in the main text there are two zeroth LLs which sit at the band edge ($E_{0^\pm} = \pm E_g \lambda/2$). The $n = 0^+$ LL only exists in satellite Dirac cones at positive wavevector along the guiding potential ($s = 1$) whilst the $n = 0^-$ LL only exists at negative wavevectors ($s = -1$). As a consequence, these zeroth LLs have half the degeneracy of the other levels meaning that if the gap were to close they would combine to a single zero-energy LL with degeneracy equal to all other levels. The wavefunctions of the zeroth LL can be written as
\begin{equation}
    \ket{\Psi_{0^\pm}(x)} = \frac{1}{\sqrt{2(1 + \lambda)l_B}} \bigg( \frac{\lambda}{\pi T} \bigg)^{\frac{1}{4}} \begin{pmatrix}
        1 + \lambda \\ -it
    \end{pmatrix} e^{-X_{0^\pm}^2/2},  
\end{equation}
for $s_\text{K} = 1$ or $-i \sigma_y \! \ket{\Psi_{0^\pm}}$ in the other graphene valley $s_\text{K} = -1$.

\bibliography{ref}

\end{document}